\documentclass[reprint,english,aps,prb,twocolumn,amsmath, amssymb, showpacs,showkeys]{revtex4}
\usepackage[T1]{fontenc}
\usepackage{subfigure}
\usepackage[latin1]{inputenc}
\usepackage{graphicx}
\usepackage{epstopdf}
\usepackage{epsfig}
\usepackage{prettyref}
\usepackage{babel}
\usepackage{color}
\makeatother

\begin{document}

\title{Spin-glass like dynamics of ferromagnetic clusters in La$_{0.75}$Ba$_{0.25}$CoO$_3$}

\author{Devendra Kumar}
\email{deveniit@gmail.com, deven@csr.res.in}
\affiliation{UGC-DAE Consortium for Scientific Research, University Campus, Khandwa Road, Indore-452001,India}

\begin{abstract}
We report the magnetization study of the compound La$_{0.75}$Ba$_{0.25}$CoO$_3$ where Ba$^{2+}$ doping is just above the critical limit for percolation of ferromagnetic clusters. The field cooled (FC) and zero field cooled (ZFC) magnetization exhibit a thermomagnetic irreversibility and the  ac susceptibility show a frequency dependent peak at the ferromagnetic ordering temperature (T$_C$$\approx$203~K) of the clusters. These features indicate about the presence of a non-equilibrium state below T$_C$. In the non-equilibrium state, the dynamic scaling of the imaginary part of ac susceptibility and the static scaling of the nonlinear susceptibility clearly establish a spin-glass like cooperative freezing of ferromagnetic clusters at 200.9(2)~K. The existence of spin-glass like freezing of ferromagnetic clusters  is further substantiated by the ZFC aging and memory experiments. We also observe certain dynamical features which are not present in a typical spin-glass, such as, initial magnetization after ZFC aging first increases and then decreases with the wait time and an imperfect recovery of relaxation in negative temperature cycling experiments. This imperfect recovery transforms to perfect recovery on concurrent field cycling. Our analysis suggests that these additional dynamical features have their origin in inter-cluster exchange interaction and cluster size distribution. The inter-cluster exchange interaction above the magnetic percolation gives a superferromagnetic state in some granular thin films but our results show the absence of typical superferromagnetic like state in La$_{0.75}$Ba$_{0.25}$CoO$_3$.
\end{abstract}



\maketitle

\section{Introduction}
Magnetically ordered materials exhibit interesting physical properties when the spin-spin correlation length is limited to the range of few nanometers.\cite{Batlle} The size of correlation length can be limited by limiting the physical size of material by forming few nanometer size single domain nanoparticles. Ensemble of such nanoparticles are observed to exhibit intriguing dynamical features, such as, superparamagnetic like thermal blocking, spin-glass like cooperative freezing, or ferromagnet like correlation between nanoparticle moments in the so called superferromagnetic state.\cite{Bedanta} The nature of physical state in the nanoparticle assembly is determined by the competition between the anisotropy energy, dipole interaction, and exchange interaction. For dilutely packed nanoparticle system a superparmagnetic state is observed, but for densely packed nanoparticle system with strong dipolar interactions, a spin-glass like cooperative freezing is reported.\cite{Bedanta,Ulrich, Sasaki} The presence of additional exchange interaction in the densely packed nanoparticle system with strong dipolar interactions can also result in a superferromagnetic state.\cite{Bedanta, Bedanta1, Frandsen, Frandsen1, Mao}

The size of correlation length can also be limited even without limiting the physical size of the system. In number of materials, e.g. in phase separated manganites (La$_{0.25}$Nd$_{0.75}$)$_{0.7}$Ca$_{0.3}$MnO$_3$,  La$_{0.7-x}$Y$_x$Ca$_{0.3}$MnO$_3$ ($0\leq x \leq 0.15$), and cobaltite La$_{1-x}$Sr$_x$CoO$_3$ ($0.18\leq x\leq 5.0$), this happens due to formation of short range ferromagnetic clusters in the paramagnetic matrix at the ferromagnetic transition.\cite{Freitas, Rivadulla, Nam} The ferromagnetic clusters in manganites are metallic and are associated with the temperature driven first-order insulator to metal transition. These clusters appear at the transition temperature, grow in number and size on lowering the temperature, and finally may undergo a spin-glass like freezing on further decreasing the temperature. The inter-cluster interaction increases on decreasing the temperature and attains its peak value at the freezing temperature.\cite{Freitas, Rivadulla} Above the magnetic percolation of ferromagnetic clusters, there is a finite possibility of exchange interaction between the neighbouring clusters in addition to the long range dipolar interaction. This additional exchange interaction, present above the magnetic percolation of Fe nanoparticles in FeAg granular thin films cause a crossover from spin-glass like state to a superferromagnetic state.\cite{Alonso}

The presence of short range ferromagnetic clusters in cobaltites with formula La$_{1-x}$A$_x$CoO$_3$ (where A is divalent ion Ba$^{2+}$ or Sr$^{2+}$) have been observed in a number of reports.\cite{Samal, Wu, Devendra1, Mandal, Kriener1, Phelan, Tong, Smith, Nam} The density of ferromagnetic clusters increases on increasing the Ba$^{2+}$ or Sr$^{2+}$ doping, and above a critical doping ($x_c$), the ferromagnetic clusters percolate. The percolation occurs at $x_c$=0.2 for Ba$^{2+}$ and $x_c$=0.18 for  Sr$^{2+}$.\cite{Samal, Wu, Kriener1, Mandal} The ferromagnetic clusters tend to retain their cluster nature and do not completely agglomerate to form a continuous phase even above the percolation.\cite{Wu, Samal, Nam, Devendra1} Below $x_c$ both the Ba$^{2+}$ and Sr$^{2+}$ doped systems exhibit a spin-glass like cooperative freezing.\cite{Mandal, Samal} Above $x_c$, ac susceptibility measurements on La$_{1-x}$Sr$_x$CoO$_3$ show some characteristics of cluster-glass dynamics while no such signature has been detected in La$_{1-x}$Ba$_x$CoO$_3$ ($x$=0.2 and 0.3).\cite{Wu,Samal, Nam, Sazonov} Recently a detailed study on La$_{0.5}$Ba$_{0.5}$CoO$_3$ has suggested the presence of interacting superparamagnetic like dynamics in the ferromagnetic clusters.\cite{Devendra1} This makes the region between $x_c$ and $x$=0.5 for Ba$^{2+}$ doping interesting as at one end we have the spin-glass like dynamics, while at the other end, we have the  interacting superparamagnetic like dynamics with surprising absence of dynamical features in between.

In this manuscript we have performed a comprehensive investigation of the magnetically ordered state of La$_{0.75}$Ba$_{0.25}$CoO$_3$ which lies just above the critical doping for percolation. The ferromagnetic clusters in La$_{0.75}$Ba$_{0.25}$CoO$_3$ will experience the long range dipolar interaction as well as the short range nearest neighbour inter-cluster exchange interaction, and therefore, are a good candidate for studying the interplay of these competing interactions. Our results show that ferromagnetic clusters in La$_{0.75}$Ba$_{0.25}$CoO$_3$ undergo a spin-glass like cooperative freezing.  In contrast to  La$_{1-x}$Sr$_x$CoO$_3$ ($0.18\leq x\leq 5.0$) and La$_{0.7-x}$Y$_x$Ca$_{0.3}$MnO$_3$ ($0\leq x \leq 0.15$), the cluster-glass transition in La$_{0.75}$Ba$_{0.25}$CoO$_3$ nearly coincides with the ferromagnetic ordering and the inter-cluster interaction is found to be unaffected of temperature.
Concurrent to the spin-glass like dynamics, we observe the signature of an additional dynamical mechanism which has been attributed to the exchange interactions between the ferromagnetic clusters. Our analysis also show that unlike FeAg granular films, the percolation of ferromagnetic clusters in La$_{0.75}$Ba$_{0.25}$CoO$_3$  do not establish a typical superferromagnetic like state in the system.

\section{Experimental Details}
Polycrystalline samples of La$_{0.75}$Ba$_{0.25}$CoO$_3$ are prepared through pyrophoric method as described in our earlier work.\cite{Devendra1} The stoichiometric ratio of high purity (99.99\%) La$_2$O$_3$, BaCoO$_3$, and Co(NO$_3$)$_2$6H$_2$O are dissolved in dilute nitric acid and triethanolamine (TEA) is added to the final solution keeping pH highly acidic. This solution is dried at 100~$^\circ$C which finally burns and yields black powder. The black powder is pelletized and heated at 1125~$^\circ$C for 12 h in air. These samples are characterized by XRD diffraction on a Bruker D8 Advance x-ray diffractometer using Cu K$\alpha$ radiation. The magnetization measurements are performed on 7T SQUID MPMS-XL (Quantum Design) and 14T PPMS-VSM (Quantum Design). The residual field in 7T SQUID MPMS-XL is set below 0.05~Oe by using the flux gate and compensation coils of ultra low field attachment before performing the zero and low field magnetization, aging, and memory experiments. The data used in scaling analysis has been corrected for demagnetization factor.

\begin{figure}[!t]
\begin{centering}
\includegraphics[width=0.6\columnwidth]{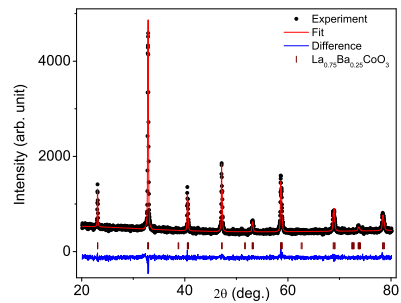}
\par\end{centering}
\caption{(Color online) X-ray diffraction pattern of La$_{0.75}$Ba$_{0.25}$CoO$_3$ at room temperature. The solid circles represents the experimental X-ray diffraction data, the red line on the experimental data exhibits the Rietveld refinement for rhombohedral R-3c structure with $\chi^2$=1.26, the short vertical lines show the Bragg peak positions, and the bottom blue line displays the difference between the experimental and calculated pattern.}\label{fig: XRD}
\end{figure}

The x-ray data has been analyzed by the Rietveld refinement method using FULLPROF software\cite{Carvajal} and the results show that the sample is single phase and crystallizes in rhombohedral structure with space group R-3c which is in agrement with the previous report.\cite{Fita} Figure~\ref{fig: XRD} displays the room temperature x-ray diffraction pattern of La$_{0.75}$Ba$_{0.25}$CoO$_3$ along with its Rietveld fit profile.  The goodness of fitting $\chi^2$ is 1.26 and the lattice parameters of the unit cell are a=5.4549(2) and c=13.3194(2). The oxygen content is determined by thermogravimetric analysis which is close to its stoichiometric value of 3.0.

\section{Results and Discussions}

\subsection{Thermomagnetic irreversibility}
\begin{figure}[!t]
\begin{centering}
\includegraphics[width=0.6\columnwidth]{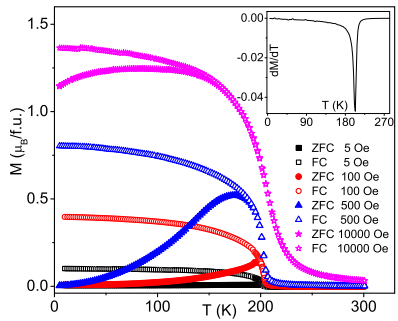}
\par\end{centering}
\caption{(Color online) Magnetization versus temperature under field cooled (FC) and zero field cooled (ZFC) protocols at various fields. The inset shows the dM/dT for 500~Oe FC data.}\label{fig: MT}
\end{figure}

Figure~\ref{fig: MT} shows the magnetization versus temperature curves at the field of 5~Oe, 100~Oe, 500~Oe, and 10000~Oe in field cooled (FC) and zero field cooled (ZFC) protocols. In FC protocol the sample is cooled to 5~K in presence of applied field and data is recorded in heating run without changing the field. In ZFC protocol the sample is cooled to 5~K in zero field, then field is applied, and data is recorded in the heating run. Both FC and ZFC magnetization curves exhibit a paramagnetic to ferromagnetic transition on cooling and the transition temperature (T$_C$) is estimated to be around 203~K by the temperature derivative of the 500~Oe FC magnetization curve (see inset of Fig.~\ref{fig: MT}). On further cooling, the ZFC magnetization curve diverges from their respective FC curve at a temperature T$_{irr}$ and exhibits a peak at a temperature T$_p$, while the FC magnetization curve continue to grow, albeit with a slower rate. On increasing the applied magnetic field, the ZFC peak flattens and T$_{irr}$ and T$_p$ shift to the lower temperatures. The existence of thermomagnetic irreversibility is
generally observed in spin-glass,\cite{Maydosh, Tiwari,Vijay}  cluster-glass,\cite{Deac, Huang} superspin-glass,\cite{Parker, Hiroi} superparamagnets,\cite{Knobel} and anisotropic ferromagnets.\cite{Devendra, Anil, Anil1, Anil2, Song, Roshkoa} The appearance of ferromagnetic ordering\cite{Tong} at T$_C$ negate the possibility of an atomic spin-glass state. The ferromagnetic state in these systems consists of small percolating ferromagnetic metallic clusters and the absence of exchange bias effect rules out the possibility of existence of a spin-glass phase at the interface of ferromagnetic-clusters.\cite{Samal, Devendra1, Mandal, Kriener1, Phelan, Tong, Smith, Gesheva} This suggests that the observed thermomagnetic irreversibility may have its origin in the dynamics of ferromagnetic-clusters.

\subsection{AC susceptibility and dynamic scaling}
\begin{figure}[!t]
\begin{centering}
\includegraphics[width=0.7\columnwidth]{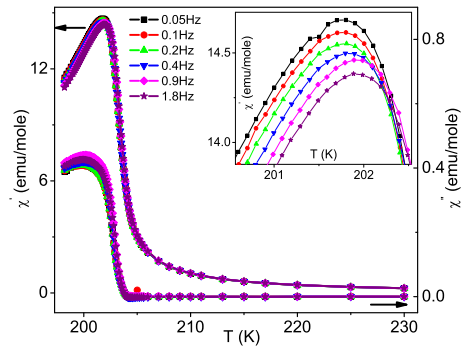}
\par\end{centering}
\caption{(Color online) Temperature dependence of real ($\chi'$) and imaginary ($\chi''$) part of ac susceptibilities at various measurement frequencies. The inset shows the expanded view of $\chi''$ close to the peak.} \label{fig: AcX}
\end{figure}

The results of ac susceptibility measurements at 0.05~Hz, 0.1~Hz, 0.2~Hz, 0.4~Hz, 0.9~Hz, and 1.8~Hz at 3~Oe ac field are displayed in Fig.~\ref{fig: AcX}. The real part of ac susceptibility ($\chi'$) exhibits a peak at the ferromagnetic transition, but as shown in the inset of Fig.~\ref{fig: AcX}, in contrast to long range ferromagnets, the peak temperature (T$_f$) increases on increasing the measurement frequency. The existence of frequency dependence in the peak position of $\chi'$ indicates that correlation length of the ferromagnetic order does not diverge at $T_C$. This frequency dependence in the peak of $\chi'$ is detectable only below 2~Hz. This is possibly due to strong contributions coming from within the ferromagnetic regions at higher frequencies. Sazonov et al.\cite{Sazonov} failed to detect the frequency dependence in $\chi'$ possibly due to higher measurement frequencies. Unlike La$_{1-x}$Sr$_x$CoO$_3$ ($0.18\leq x\leq 5.0$),\cite{Nam, Wu} we do not observe any secondary peak or hump in $\chi''$. The frequency dependence in the peak of $\chi'$ generally manifests in the non-equilibrium magnetic states, such as spin-glass, cluster-glass, superspin-glass, and super-paramagnets, and the parameter $\Phi=\Delta T_f/(T_f\Delta$log$_{10}f)$ which quantify this dependence is observed to be around (0.02-0.005) for spin-glass, cluster-glass, superspin-glass, and interacting superparamagnets and around (0.1-0.3) for non-interacting superparamagnets.\cite{Goya, Cong, Thakur, Toro, Dormann} The data shown in the inset of Fig.~\ref{fig: AcX} gives $\Phi$ around 0.0005. Since the possibility of an atomic spin-glass state or a spin-glass phase at the interface of ferromagnetic clusters have been already ruled out, the observed frequency dependence can be because of spin-glass like freezing or superparamagnet like thermal blocking of interacting ferromagnetic clusters.

\begin{figure} [!t]
\begin{centering}
\includegraphics[width=1.0\columnwidth]{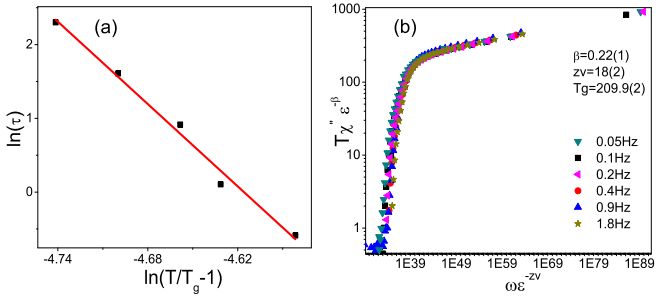}
\par\end{centering}
\caption{(Color online) (a) Frequency dependence of 
T$_f$ plotted as ln$\tau$ versus ln$\epsilon$. The straight line is the least square fitting of equation~\ref{eq:acsusp}. (b) Dynamic scaling of $\chi''(\omega,T)$ on the basis of equation~\ref{eq:acsusp1}.  } \label{fig:acscaling}
\end{figure}

If the observed frequency dependence of T$_f$ is due to critical slowing down of fluctuating clusters, as in the case of spin-glass transition, the spin-cluster correlation length ($\xi$) should diverge as $\xi\propto\epsilon^{-\nu}$ on approaching T$_g$ from T>T$_g$. Here T$_g$ is the cluster-glass transition temperature at the zero field dc limit, $\epsilon$=$(T-T_g)/T_g$ is the reduced temperature, and $\nu$ is the static critical exponent. The relaxation time ($\tau$) is related to correlation length ($\xi$) as $\tau\propto\xi^{z}$ where $z$ is the dynamic critical exponent. Thus for a cluster-glass transition, the relation for relaxation time $\tau$ (corresponding to a given measurement frequency $f$, $\tau$=$f^{-1}$) can be written as
 \begin{equation}
\tau=\tau_0(T/T_g-1)^{-z\nu},\label{eq:acsusp}
\end{equation}
 where $\tau_0$ is the spin flipping time of fluctuating spin-clusters. For a given measurement frequency $f$,  the peak temperature of corresponding $\chi'(f)$,  i.e. T$_f$, is associated with the cluster-glass transition temperature. The imaginary part of ac susceptibility ($\chi''(f)$) also exhibits a frequency dependent peak, and in this case, the inflection point in $\chi''(f)$ is identified as the cluster-glass transition temperature (T$_f$). Through T$_f$ calculated from $\chi'(f)$ and $\chi''(f)$ is expected to show qualitatively similar features, frequency dependence in $\chi''(f)$ is more pronounced, and therefore estimation of T$_f$ from inflection point of $\chi''(f)$ is relatively more accurate. For 0.05~Hz, the error in estimation of T$_f$ is large and so it has been left out from the fitting process. Fig~\ref{fig:acscaling}~(a) shows the fitting of equation~\ref{eq:acsusp} to $\tau$ versus T$_f$ data following the procedure described in Ref.~\onlinecite{Devendra1}. The data fits well with $\tau_0\sim10^{-38}$s, T$_g$=200.9(1)K, and $z\nu$=18(1).  While the good fitting of equation~\ref{eq:acsusp} to the data suggests the possibility of spin-glass like phase transition, the value of fitting parameter $\tau_0$ is unphysical and $z\nu$ is outside the range of canonical spin-glass(4-10)\cite{Souletie, Fischer} but is close to the values reported in cluster or superspin -glass(10-15).\cite{Parker, Svanidze, Vijayanandhini, Hiroi, Anand} The anomaly in $\tau_0$ can be due to error in estimation of fitting parameter caused by limited span of frequency range.

In critical slowing down description of spin-glass phase transition, $\chi''(\omega,T)$ should behave according to the dynamic scaling equation proposed by Geschwind et al.\cite{Geschwind}
\begin{equation}
T\chi''(\omega,T)\epsilon^{-\beta}=g(\omega\epsilon^{-z\nu}),\label{eq:acsusp1}
\end{equation}
where $\omega=2\pi f$, $\beta$ is the critical exponent corresponding to the order parameter, and $g$ is a universal function of its argument.
If the frequency dependence in $\chi''$ is indeed due to spin-glass like phase transition, then $\chi''$ curves of different frequency should collapse on a single universal curve ($g$) for the proper value of critical exponents $\beta$, $z\nu$ and transition temperature T$_g$. As shown in Fig.~\ref{fig:acscaling}~(b), a  nearly perfect collapse of the data over two decades of frequency is obtained for T$_g$=209.9(2)~K, $\beta$=0.22(1) and $z\nu$=18(2) which confirms the presence of spin-glass like phase transition in the system. The value of the parameters $z\nu$ and T$_g$ are in agrement with the values obtained from equation~\ref{eq:acsusp}.

\subsection{Static scaling of nonlinear susceptibility}
\begin{figure} [!t]
\begin{centering}
\includegraphics[width=0.5\columnwidth]{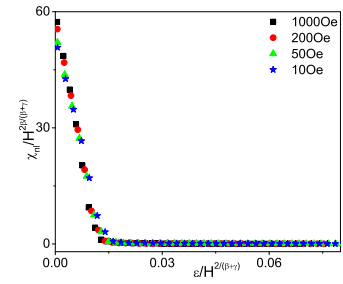}
\par\end{centering}
\caption{(Color online) Scaling plot of the nonlinear susceptibility above T$_g$ based on the equation~\ref{eq:static scaling}.} \label{fig:static scaling}
\end{figure}

The magnetization ($M$) at the uniform applied field $H$ can be expanded in terms of nonlinear susceptibilities as
\begin{equation}
M(H)=M_0 + \chi_1H + \chi_2H^2 + \chi_3H^3+....,\label{eq:Mag}
\end{equation}
where $M_0$ is the spontaneous magnetization, $\chi_1$ is the linear susceptibility, and $\chi_2$, $\chi_3$.. are nonlinear susceptibilities. For an atomic spin-glass, $M_0$ and the coefficients of even power of $H$ i.e $\chi_2$, $\chi_4$.. are zero while coefficients of the odd power of $H$ i.e $\chi_3$, $\chi_5$.. diverges as $T$ approaches $T_g$ in the critical regime.\cite{Suzuki, Wada, Fujiki} For cluster-glass, if the nonlinear response of isolated ferromagnetic clusters is small, coefficients of the odd power of $H$ will also diverge in the critical regime similar to that of atomic spin-glass.\cite{TJonsson} The overall nonlinear susceptibility $\chi_{NL}$ which diverges in the critical temperature regime in a spin-glass system can be written as

\begin{equation}
\chi_{NL}=\chi_1-M/H= \chi_3H^2+\chi_5H^4+.... .\label{eq:dc scaling}
\end{equation}
The phenomenological theory of spin-glass by Suziki predicts that the $\chi_{NL}$ should follow the static scaling relation\cite{Suzuki}
\begin{subequations}
\begin{align}
\chi_{NL}&=\epsilon^{\beta}F(H^2/\epsilon^{\beta+\gamma})\label{second}\\
\text{or}\;\;\;\;
\chi_{NL}&=H^{2\beta/(\beta+\gamma)}G(H^2/\epsilon^{\beta+\gamma}),\label{third}
\end{align}
\end{subequations}
where $\epsilon$=(T-T$_g$)/T$_g$ is the reduced temperature, $\beta$ is the critical exponent for spin-glass order parameter, $\gamma$ is the critical exponent for spin-glass susceptibility, and $F(x)$ and $G(x)$ are the scaling functions. The scaling is achieved by plotting $\chi_{NL}/H^{2\beta/(\beta+\gamma)}$ versus  $H^2/\epsilon^{\beta+\gamma}$ for $\chi_{NL}$ at different fields and varying the parameters $T_g$, $\beta$ and $\gamma$ such that all data collapse on a master curve. In the limit of $\epsilon\rightarrow$0 the abscissa and ordinate have a span of many decades, and therefore, are plotted on log scales. The log scale plotting gives equal weightage to all data points irrespective of their accuracy which sometimes hides the departure from the good scaling.\cite{Geschwind1, Tiwari} To test the scaling equation in a better way, Geschwind et al.\cite{Geschwind1} have rewritten the scaling equation such that the argument of scaling function is linear in $\epsilon$
\begin{equation}
\chi_{NL}=H^{2\beta/(\beta+\gamma)}
\tilde{G}(\epsilon/H^{2/{(\beta+\gamma)}}).\label{eq:static scaling}
\end{equation}
The contribution of nonlinear susceptibilities to magnetization diminishes as $H$ approaches to zero, and in this limit, the magnetization above T$_g$ gives a reasonable approximation of $\chi_1$. We have used the magnetization at 0.5~Oe as the approximate value of $\chi_1$. Figure~\ref{fig:static scaling} shows the scaling plot obtained using equation~\ref{eq:static scaling} for the data at 10~Oe, 50~Oe, 200~Oe, and 1000~Oe. The data taken at four different fields collapse best on a master curve for $\beta$=0.22, $\gamma$=40, and $T_g$=200.9~K. The reasonable scaling of the $\chi_{NL}$ by equation~\ref{eq:static scaling} supports the occurrence of a spin-glass like phase transition at $T_g$. The values of parameter $\beta$ and $T_g$ are in agrement with that of the dynamic scaling. The parameter $\beta$ lies in the range of cluster or superspin -glass while parameter $\gamma$ is large in comparison to typical values observed in spin-glass, cluster-glass, or superspin-glass.\cite{TJonsson, RMathieu, Rivadulla, Fischer}

\subsection{Aging and memory experiments}
\label{AgingMemory}
\begin{figure} [!t]
\begin{centering}
\includegraphics[width=0.9\columnwidth]{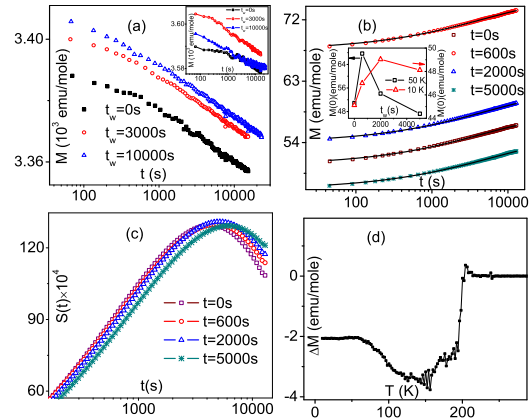}
\par\end{centering}
\caption{(Color online) (a) Wait time dependence of decay of thermoremanent magnetization (TRM) at 80~K. The inset shows the wait time dependence of TRM at 50~K. (b) Wait time dependence of ZFC magnetization at 50~K. The solid line is the fit of equation~\ref{eq:stretchexp} to data. The inset shows the wait time dependence of $M(0)$ at 50~K and 10~K. (c) $S(t)$=$(1/H)(dM(t)/$ln$t$) for different wait times at 50~K. (d) The temperature dependence of $\Delta M$=$M_{ZFC}^{mem}$-$M_{ZFC}^{ref}$. $M_{ZFC}^{mem}$ has an intermediate stop of 10$^4$~s at 130~K in the cooling run. } \label{fig:agingmemory}
\end{figure}

The existence of a spin-glass like phase transition in La$_{0.75}$Ba$_{0.25}$CoO$_3$ is also investigated through aging and memory experiments. In FC (ZFC) aging experiments, the system is cooled from 250~K to temperature of aging (T$_a$) in presence of field $H_a$ (zero filed), and at T$_a$ after isothermal waiting of $t_w$~s, the field is switched off (field $H_a$ is applied) and the magnetization is recorded as a function of time.  The results of field cooled aging experiments at 80~K and 50~K for $H_a$=500~Oe are presented in Fig.~\ref{fig:agingmemory}~(a) and its inset respectively. The decay of thermoremanent  magnetization exhibits the effect of wait time dependence which has been reported to occur in superparamagnets with distribution of anisotropy energy barriers as well as in systems with a spin-glass like transition.\cite{Sasaki} The value of initial magnetization observed after switching off the field at 80~K increases on increasing the wait time similar to that of superparamagnets or spin-glass but exhibits a complex behavior at 50~K. The aging experiments performed in the ZFC protocol show a distinct behavior in superparamagnet and spin-glass systems. In superparamagnets the wait time dependence in ZFC aging is either absent or significantly small (when several competing sources of anisotropy are present) in comparison to the wait time dependence of FC aging, while in spin-glass systems, a strong wait time dependence is observed both in case of FC as well as ZFC aging.\cite{Sasaki, Vijay} Figure~\ref{fig:agingmemory}~(b) exhibits the results of ZFC aging experiments at T$_a$=50~K and H$_a$=100~Oe. The presence of strong wait time dependence in ZFC aging as in the case of FC aging reconfirms the presence of spin-glass like transition in the system. The observed time dependence in magnetization is fitted with stretched exponential decay
 \begin{equation}
M(t)=M_0-M_gexp(-(t/\tau)^{\beta}),\label{eq:stretchexp}
\end{equation}
where $M_0$ is the contribution of intrinsic ferromagnetic component, $M_g$ is the initial magnetization of the glassy component, $\tau$ is the time constant, and $\beta$ represents the distribution of energy barriers with $0\leq\beta\leq1$ for spin-glass. The equation~\ref{eq:stretchexp} fits well to the time dependent magnetization data of Fig.~\ref{fig:agingmemory}~(b) and the fitting parameters are shown in Table~\ref{tab:fitting-parameters}. The parameters $M_g$ and $\tau$ increase while $\beta$ decreases on increasing the wait time. In spin-glass systems, the magnetization at t=0 i.e. M(0) is observed to decrease on increasing the wait time ($t_w$),\cite{Sasaki, Vijay, Lundgren}  but in La$_{0.75}$Ba$_{0.25}$CoO$_3$, it first increases and then decreases. The ZFC aging experiments performed at 10~K (not shown here) also give similar results. A comparison of wait time dependence of M(0) at 50~K and 10~K is shown in inset of Fig.~\ref{fig:agingmemory}~(b). A similar behavior in M(0) is also observed in ZFC aging of ferromagnetic La$_{0.8}$Ca$_{0.2}$MnO$_3$ (18~nm) and antiferromagnetic La$_{0.2}$Ca$_{0.8}$MnO$_3$ (15~nm) nanoparticles.\cite{Markovich, Markovich1} The initial increase in M(0) with $t_w$ suggests the presence of an additional dynamical mechanism along with the spin-glass like freezing. In La$_{0.8}$Ca$_{0.2}$MnO$_3$ nanoparticles, the additional dynamic features have been claimed to be coming from the development of a superferromagnetic (SFM) like state i.e. development of ferromagnetic like correlation among the superspins.\cite{Markovich}

For spin-glass, Lundgren et. al. have suggested that the quantity $S(t)$=$(1/H)(dM(t)/$ln$t$) is proportional to the spectral density of relaxation times ($g(\tau)$), and therefore, $S(t)$ versus $t$ plots give an estimate of $g_{tw}(\tau=t)$ for given  $t_w$.  As $t_w$ increases, $g_{tw}(\tau)$ shifts towards longer relaxation times and peaks around $t_w$.\cite{Lundgren, Lundgren1, Vincent} Figure~\ref{fig:agingmemory}~(c) show the $S(t)$ versus ln$t$ curves for various $t_w$. As expected in a spin-glass system, we get a peak in $S(t)$ at $t^{eff}_w$ which shifts to higher $t$ on increasing $t_w$ but the order of shift is relatively small in comparison to atomic spin-glass. In atomic spin-glass $t^{eff}_w$$\approx$$t_w$, while in our case, t$^{eff}_w$>$t_w$ and show a weak $t_w$ dependence. $t^{eff}_w$ varies from  4760~s to 6144~s on changing $t_w$ from 600~s to 5000~s. The observed $t_w$ dependence of $t^{eff}_w$ resembles with that reported for La$_{0.7-x}$Y$_x$Ca$_{0.3}$MnO$_3$ ($0\leq x \leq 0.15$) manganites which exhibit a cluster-glass behavior and La$_{0.8}$Ca$_{0.5}$MnO$_3$ (18~nm) nanoparticles which exhibit a superspin-glass like behavior.\cite{Freitas, Markovich}

\begin{table}[!t]
\begin{centering}
\begin{tabular}{|c|c|c|c|c|c|c|}
\hline\  $t_w$ (s) &  $M_0$ & $M_g$ & $\tau$$\times$10$^3$  & $\beta$ &$\chi^{2}/DOF$
& $R^{2}$  \tabularnewline \hline\ 0 & 57.73(8) & 7.0(1) &
4.3(2)& 0.502(8) & 0.00142 & 0.99920 \tabularnewline \hline\ 6$\times$10$^2$ & 74.77(9) & 7.2(1) &
4.8(2)& 0.488(8) & 0.00139 & 0.99921 \tabularnewline
\hline\ 2$\times$10$^3$  & 61.26(9) & 7.3(1) & 5.2(2)& 0.488(7) & 0.00103 & 0.99941
\tabularnewline \hline\ 5$\times$10$^3$  & 54.5(1) & 7.4(1) &
6.1(3)& 0.476(7) & 0.00027 & 0.99982 \tabularnewline \hline
\end{tabular}
\par\end{centering}
\caption{Fit parameters obtained from the fitting of equation~\ref{eq:stretchexp} to the magnetization data of Fig.~\ref{fig:agingmemory}~(b). \label{tab:fitting-parameters}}
\end{table}

The results of ZFC memory experiments are displayed in Fig.~\ref{fig:agingmemory}~(d). In ZFC memory experiment, the system is cooled in zero field from 300~K to 10~K with an intermediate stop of 10$^4$~s at 130~K, and at 10~K, 100~Oe field is applied and the magnetization ($M_{ZFC}^{mem}$) is recorded in the heating run. The reference ZFC magnetization ($M_{ZFC}^{ref}$) is also recorded under same protocol but without an intermediate stop. The difference in magnetization i.e. $\Delta M$=$M_{ZFC}^{mem}$-$M_{ZFC}^{ref}$ versus temperature exhibits a broad dip around 140~K which signals that the system remembers its aging at the intermediate stop during the cooling run. The memory in ZFC protocol is only observed in systems undergoing a spin-glass like cooperative freezing, and therefore, the observation of ZFC memory in Fig.~\ref{fig:agingmemory}~(d) confirms the existence of a spin-glass like state in La$_{0.75}$Ba$_{0.25}$CoO$_3$.\cite{Sasaki, Bandyopadhyay, Mathieu, Vijay} Here we note that in contrast to atomic spin-glass, the observed dip in ZFC memory is broad and does not exhibit a complete rejuvenation. Such a behavior is generally observed in superspin-glass or cluster-glass where the microscopic flipping time of the fluctuating magnetic entities (superspins) is much longer than that of  atomic spin-glass, and therefore, the observation time in units of microscopic flipping time is relatively shorter in comparison of atomic spin-glass.\cite{Markovich, Markovich1, Jonsson, Sankar} Because of this, the length scales probed during the experimental time scale are shorter and the condition for rejuvenation, i.e. the length scales probed during the experimental time scale are larger than the so called overlap length, is not satisfied.\cite{Jonsson} The absence of complete rejuvenation in ZFC memory may also occur if the probing field is strong enough to perturb the intrinsic non-equilibrium dynamics of the spin-glass system.

\subsection{Temperature-field cycling and relaxation}
\begin{figure} [!t]
\begin{centering}
\includegraphics[width=0.9\columnwidth]{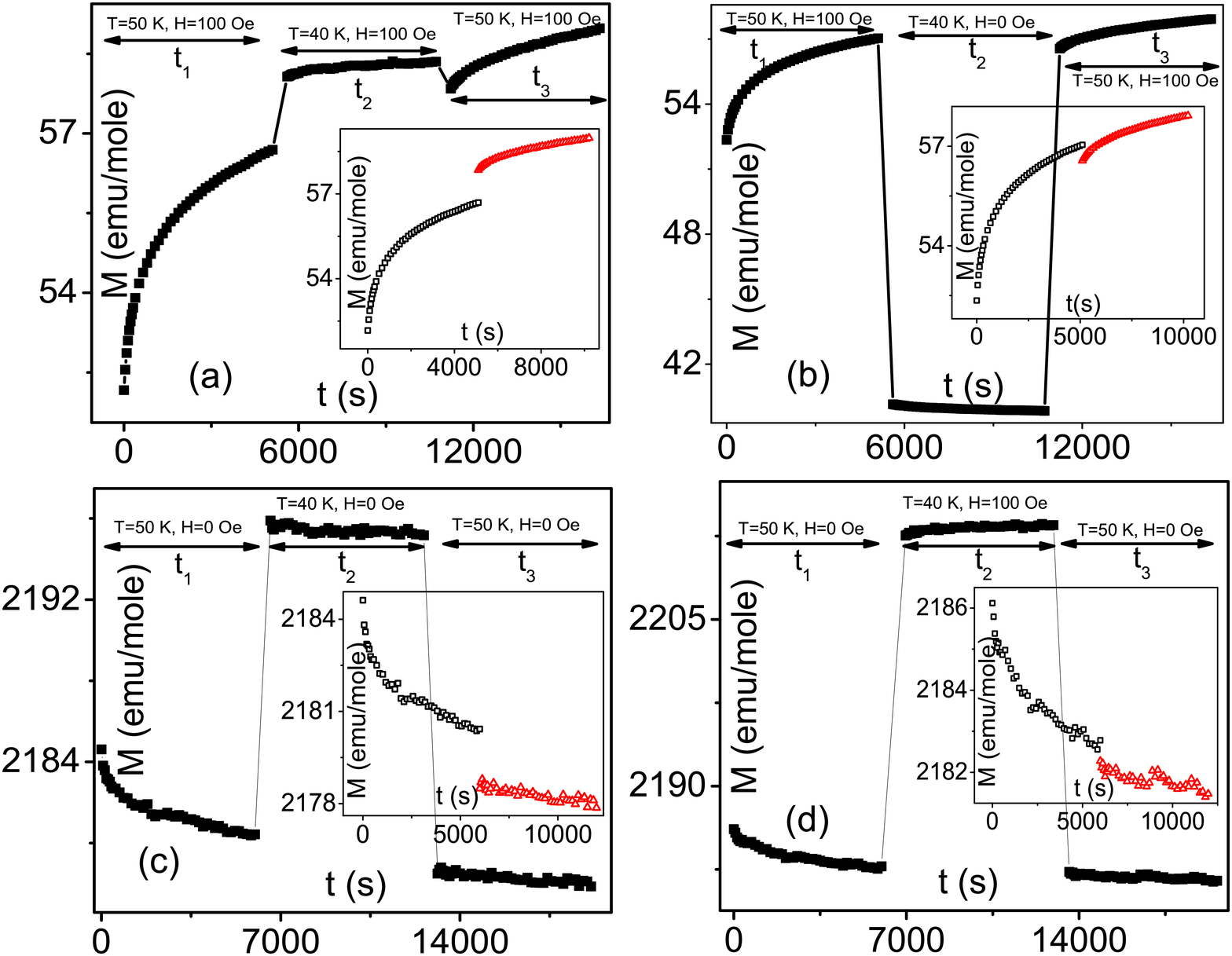}
\par\end{centering}
\caption{(Color online) Magnetization relaxation with an intermediate negative temperature cycling (a) in ZFC protocol without field change, (b) in ZFC protocol  with field change, (c) in FC protocol without field change, and (d) in FC protocol with field change. Insets show the relaxation at $t_1$ and $t_3$ only (after adjusting the time scale for $t_2$) for the respective measurement protocols.} \label{fig:AgingRej}
\end{figure}

The relaxation experiments discussed in previous section are complemented with an intermediate negative
thermal cycling with and without field change. Figure.~\ref{fig:AgingRej}~(a) shows the result of temperature cycling without field change under ZFC protocol. Here the sample is cooled to 50~K in zero field, then 100~Oe field is applied and the magnetization is recorded as a function of time for $t_1$~s. Then keeping the field constant, the temperature is changed to 40~K and magnetization is recorded for $t_2$~s. Thereafter temperature is again brought to 50~K (without changing the field) and magnetization is taken for $t_3$~s. According to Sun et al., for superspin or cluster -glass, the relaxation during $t_3$ should be the continuation of the relaxation at $t_1$.\cite{Sun, Bandyopadhyay, Vijay} The inset of Fig.~\ref{fig:AgingRej} (a) shows the relaxation at $t_1$ and $t_3$. The relaxation during $t_3$ has the same functional form as during $t_1$ but is shifted upward. Figure.~\ref{fig:AgingRej}~(b) exhibits the relaxation under similar protocol but with a zero field between $t_1$ and $t_3$. As clear from the inset of Fig.~\ref{fig:AgingRej}~(b), the relaxation at $t_3$ is the continuation of $t_1$ without any apparent shift in magnetization as expected from a typical superspin or cluster -glass. This indicates that a concurrent field cycling destroys the mechanism responsible for the observed additional upward shift in magnetization during the negative thermal cycling. The results of similar set of negative thermal cycling experiments performed with and without field cycling under FC protocol are shown in Fig.~\ref{fig:AgingRej} (c) and (d) respectively. The insets show the relaxation during $t_1$ and $t_3$. When field remains zero during the temperature cycling, the relaxation at $t_3$ has the same functional form as $t_1$ but starts with a lower initial value. A concurrent field cycling along with temperature cycling (0Oe, 50K, $t_1$s, - 100Oe, 40K, $t_2$s, - 0Oe, 50K, $t_3$s) nearly removes this shift, and in this case, relaxation during $t_3$ is a continuation of relaxation during $t_1$ similar to that observed in superspin or cluster -glass. The results of negative temperature cycling without the field change in ZFC and FC protocols support the finding of ZFC aging experiments (discussed in subsection~\ref{AgingMemory}) that there exists an additional  relaxation mechanism apart from the usual cluster-glass relaxation. This additional relaxation mechanism seems to enhance the effect of cluster-glass relaxation, i.e. it contributes positively to magnetization when spin-clusters align during the relaxation process (ZFC aging) but contributes negatively to magnetization when spin-clusters randomize during the relaxation (FC aging). The field cycling probably blocks the continuous contribution of this additional relaxation mechanism to magnetization during negative temperature cycling, and thus, leave us with a normal cluster-glass like relaxation behavior.

\section{Further discussions}
The experimental results discussed in the previous sections clearly establish the existence of a spin-glass like phase transition along with an additional concurrent dynamics in the so called cluster-ferromagnetic state of La$_{0.75}$Ba$_{0.25}$CoO$_3$. The spin-glass like behavior may arise due to cooperative freezing of ferromagnetic clusters or it can be due to coexistence of a spin-glass phase with the ferromagnetic-clusters. The absence of exchange bias effect suggests the lack of ferromagnet spin-glass interface. This along with the values of critical exponents $zv$ and $\beta$, the wait time dependence of $S(t)$, and the lack of complete rejuvenation in ZFC memory indicate that the spin-glass like behavior is coming from the cooperative freezing of ferromagnetic-clusters instead of atomic spins.

The existence of spin-glass like dynamics in assembly of ferromagnetic clusters require (a) dense packing of ferromagnetic clusters  with random orientation of anisotropy axes, and (b) strong inter-cluster dipolar interactions. In La$_{0.75}$Ba$_{0.25}$CoO$_3$, the ferromagnetic clusters with randomly oriented anisotropy axes percolates,\cite{Mandal, Kriener1, Phelan, Tong, Smith} and so, there is also a reasonable possibility of exchange interaction between the neighboring ferromagnetic clusters. The exchange interaction between the ferromagnetic clusters can be because of exchange bridges between the surface atoms of neighbouring ferromagnetic clusters or can be due to tunneling exchange coupling between the neighbouring metallic ferromagnetic clusters.\cite{Hansen, Frandsen, Frandsen1, Kondratyev}  The competing dipolar and inter-cluster exchange interaction in a disordered random anisotropy system can lead to a superferromagnetic  state.\cite{Hansen, Kechrakos, Bedanta, Bedanta1, Alonso} The superferromagnetic state exhibits dynamic features as in the cluster (or superspin)-glass, but in contrast to cluster-glass which has a zero thermoremanence magnetization in the limit of $t\rightarrow$$\infty$, superferromagnet has a finite remanence.\cite{Chamberlin, Chen, Chen1}

\begin{figure} [!t]
\begin{centering}
\includegraphics[width=0.5\columnwidth]{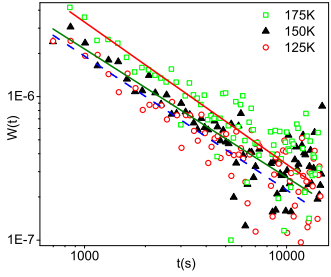}
\par\end{centering}
\caption{(Color online) Relaxation rate versus time on log scale at 175~K, 150~K, and 125~K. The straight line show the fitting of equation~\ref{eq:TRMpowerlaw} to  respective data.} \label{fig:interaction}
\end{figure}

The decay of relaxation rate ($W(t)$) of thermoremanent magnetization ($m(t)$) can be used to distinguish the cluster (or superspin)-glass dynamics from that of superferromagnet. According to Monte Carlo simulations of Ulrich et al., the $W(t)$ of an assembly of 
nano-particles with dipole interaction decays as a universal power law after some crossover time $t_0$\cite{Ulrich}
 \begin{equation}
W(t)=\frac{d}{dt}\text{ln} m(t)=At^{-n},\label{eq:TRMpowerlaw}
\end{equation}
where exponent $n$ depends on the packing density of nano-particles and  $A$ is a temperature dependent constant. Depending on the value of exponent $n$, the $m(t)$ decays as stretched exponential $m(t)$=$m_0$exp($-(t/\tau)^{1-n})$ (for $n$<1), power law $m(t)$=$m_1t^{-A}$  (for $n$=1), or power law with finite remanence $m(t)$=$m_{\infty}+m_1t^{1-n}$ (for $n$>1). The decay with exponent $n$<1 is associated with dilute systems, $n$=1 is associated with cluster or superspin -glass, and  $n$>1 is associated with superferromagnets.\cite{Ulrich, Chen, Chen1} These theoretical predictions have been substantiated by relaxation measurements on granular multilayers, magnetic clusters, and nanoparticle assemblies.\cite{Chen, Chen1, Patra, Rivadulla, Tang1, De, Wang, Karmakar} Experimentally $n\approx1$ ($n\ngtr1$) is observed for cluster or superspin -glass while $n>1$ for superferromagnets. Mao et al. have performed mean field calculations including both the dipolar and exchange interactions among the nanoparticles and their results show that $n\approx1$ (with superspin-glass behavior) for small exchange interactions which smoothly transforms to $n$>1 (with superferromagnetic behavior) on gradually enhancing the strength of exchange interaction keeping dipolar interaction intact.\cite{Mao}

Figure \ref{fig:interaction} shows the $W(t)$ versus time on the log-log scale at 175~K, 150~K, and 125~K. Here the system is cooled from 300~K to respective relaxation temperatures under a field of 100~Oe, and after temperature stabilization, the field is switched off and the thermoremanent magnetization is recorded as a function of time. The fitting of equation \ref{eq:TRMpowerlaw} to the $W(t)$ versus time data gives $n$=0.99(7) at 175~K, $n$=0.89(4) at 150~K and $n$=0.93(3) at 125~K. The $n$ values are close to one (from lower side) supporting the cluster-glass dynamics. We have also fitted the time dependence of thermoremanent magnetization with stretched exponential, power law, and power law with finite remanence and only the power law function fits with reasonable parameters and error values. The power law with finite remanence fit gives huge error in fitting parameters. This along with $n$<1 rules out the possibility of superferromagnet like dynamics and suggests that inter-cluster exchange interaction in La$_{0.75}$Ba$_{0.25}$CoO$_3$, if present, are relatively weak in comparison to the dipolar interaction.  We also note that the value of exponent $n$ remains nearly constant on changing the relaxation temperature from 125~K (0.6255T$_g$) to 175~K (0.875T$_g$) which indicates that the inter-cluster interaction remains nearly constant on approaching T$_g$.
This behaviour is quite different from other phase separated mangnaites and cobaltites where $n$ increases on approaching $T_g$ due to increase in inter-cluster dipolar interactions which has been attributed to enhancement in cluster density.\cite{Rivadulla, Tang1}

The presence of additional dynamic features in the cluster-glass state can be understood by taking into account the competition between the dipolar and exchange interactions. The energy $E_i$ of a cluster $i$ in an ensemble of ferromagnetic clusters can be written as
\begin{multline}
E_i=-KV_i(\boldsymbol{\hat{\mu}_i.\hat{n}_i})^2 + \sum\limits_{j}\frac{(\boldsymbol{\mu_i.\mu_j})-3(\boldsymbol{\mu_i.\hat{r}_{ij}})(\boldsymbol{\mu_j.\hat{r}_{ij}})}{r_{ij}^3}\\
-\sum\limits_{j(nn)}J_{ij}\boldsymbol{\mu_i.\mu_j}-\boldsymbol{\mu_i.H},\label{eq:TRM}
\end{multline}
where $K$ is the anisotropy constant, $\boldsymbol{\hat{n}_i}$ is the unit vector along the easy axis,  $V_i$ is the volume, $\boldsymbol{\mu_i}$ is the moment of $i$th cluster, $\boldsymbol{r_{ij}}$ is the distance between the $i$th and $j$th cluster and $\boldsymbol{H}$ is the applied field. $J_{ij}$ represents the exchange coupling between the nearest neighbour $i$ and $j$. The anisotropy energy favours the alignment of cluster moment along the easy axis $\boldsymbol{\hat{n}_i}$ which may vary randomly from cluster to cluster. The first term of dipolar energy favours antiferromagnetic coupling while the second term attempt to align these clusters randomly. The anisotropy energy and the dipolar interaction in a densely packed cluster system give a spin-glass like cooperative dynamics.\cite{Ulrich, Mao} The exchange energy favours the alignment of nearest neighbours while Zeeman energy favours the alignment of all the cluster moments along the field direction. The anisotropy energy gives two equal energy minima, one along and the other opposite to $\boldsymbol{\hat{n}_i}$. The dipolar energy, the exchange energy, and the Zeeman energy may lower the energy of one of these minima and the cluster moment will transform to the low energy state by thermal activation.  For weak exchange interaction and small field, the dominant anisotropy and dipolar energy terms cause a cluster-glass like dynamics.\cite{Mao} The ensemble of ferromagnetic clusters in our system will have a cluster size distribution. The small clusters have a larger surface to volume ratio, and therefore, the exchange interaction may be more significant for small clusters embedded between the larger clusters. The dynamics of these small clusters will strongly depend on the alignment of their neighboring moments.  The additional dynamical behavior observed in the aging and negative temperature cycling experiments can be possibly understood on the basis of the above picture as follows:
\begin{enumerate}
\item	After a zero field quench from above T$_g$, during zero field aging, the small embedded clusters will try to align according to their big neighbors because of significant exchange interaction while the big ones will relax according to the spin-glass model. The initial enhancement in magnetization is possibly a result of thermally activated alignment of small clusters which is slowly overcome by the dominant spin-glass like relaxation of larger clusters.

\item The applied field lowers the energy of easy axis direction pointing along the field. If this now becomes the low energy state, the activated alignment of cluster moments along the field direction will enhance the magnetization with time. Switching off the field favors the restoring of spin-glass like ordering due to prevalent dipole interaction. This causes magnetization to decay with time. Because of the hierarchical nature of dynamics in spin-glass, the set of energy barriers relaxed at temperature T are different from that of T-$\Delta$T. When the temperature is lowered to T-$\Delta$T after a relaxation at T, the contribution to relaxation at T-$\Delta$T comes from (i) the spin-glass like relaxation of active clusters (at T-$\Delta$T) and the exchange induced activated alignment of their small neighbors and (ii) exchange induced activated alignment of small clusters by the frozen neighbours which were active at T.
\item If the field is not changed during thermal cycling then exchange induced activated alignment of small clusters in the above  process (i) and (ii) adds up. This gives an additional upward shift (downward shift) in magnetization for ZFC protocol (for FC protocol) when the temperature is brought back to T. Switching off (on) the field during relaxation in ZFC protocol (FC protocol) at T-$\Delta$T weakens (ii) because now Zeeman energy does not support (oppose) the alignment effort of exchange energy. Additionally, the contribution from alignment of small cluster in (i) will be opposite of (ii). Because of the opposite sign in contribution from (i) and (ii), they tends to cancel each other and therefore there is no significant shift in magnetization when the temperature is raised back to T.
\end{enumerate}

\section{Conclusions}
We have performed a detailed investigation of La$_{0.75}$Ba$_{0.25}$CoO$_3$ which lies just above the critical doping for percolation of ferromagnetic clusters. Our results show an irreversibility in FC-ZFC magnetization and a frequency dependent peak in ac susceptibility which coincides with the ferromagnetic ordering (T$_C\approx203~K$) of the clusters. The contribution from magnetic ordering within the clusters mask the frequency dependence of ac susceptibility above 2~Hz. The FC-ZFC irreversibility and frequency dependence in the peak of ac susceptibility indicate about the existence of a non-equilibrium state below T$_C$. The dynamic scaling of the imaginary part of ac susceptibility, the static scaling of the nonlinear susceptibility, and the ZFC aging and memory experiments give conclusive evidence of the spin-glass like cooperative freezing of ferromagnetic clusters (T$_g$=200.9(2)~K) in the non-equilibrium state.

The results of ZFC aging experiments in the non-equilibrium state indicate about the existence of an additional dynamical mechanism apart from the typical spin-glass dynamics. The presence of this additional dynamical mechanism is further substantiated by relaxation under negative temperature cycling. Our analysis show that this additional dynamical mechanism possibly have its origin in the inter-cluster exchange interaction and cluster size distribution. The inter-cluster exchange interactions can create a superferromagnetic state in the ensemble of densely packed ferromagnetic clusters. The decay of relaxation rate of thermoremanent magnetization and the decay of thermoremanent magnetization suggest the absence of  typical superferromagnetic state in La$_{0.75}$Ba$_{0.25}$CoO$_3$.

\subsection{Acknowledgement}
We are thankful to A. Banerjee for support and discussions. We also acknowledge M. Gupta for XRD measurements.

\end{document}